# Expanded Comment on:
# Optical Properties of Fluid Hydrogen at the Transition to a Conducting State

Isaac F. Silvera[1], Rachel Husband[1], Ashkan Salamat[1,2], and Mohamed Zaghoo[1]
[1]Lyman Laboratory of Physics, Harvard University, Cambridge, MA 02421
[2]University of Nevada, Las Vegas, NV, 89154

In a recent paper McWilliams et al (**MDMG**) [1] measured transmittance, **Tr**, of high pressure-high temperature hydrogen, heating to temperatures as high as 6000 K. They claim that hydrogen is a semiconductor or semi-metal at their conditions and there is no evidence of a first-order phase transition to metallic hydrogen (**MH**). They laser heat an absorber embedded in hydrogen. The absorber has a central hole and they measure the transmittance through the hole. Their claim is based on a weak transmitted light signal interpreted as absorption in hydrogen that increases with increasing wavelength; they support this by a finite element analysis (**FEA**) and a nine-parameter fit to their data. They use a spatial filter [2] to restrict the light that falls on the detector to that coming from the hole in the absorber. A spatial filter uses a lens to form a real magnified image of the sample in a focal plane; a small aperture in this plane is used to select the light from the absorber hole. The claim by MDMG is in contradiction with direct measurements of Tr and reflectance **R** of pressurized and heated dense hydrogen by Zaghoo, Salamat, and Silvera (**ZSS**) [3]. **ZSS** provided strong evidence for an abrupt phase transition to a reflecting metallic phase of hydrogen when it is heated above a certain temperature identified as the phase transition temperature at the given pressure. They measured the Pressure-Temperature (**P-T**) phase line and optical reflectance of MH, R~0.5. Shock experiments on deuterium measured R ~0.5 [4,5]. There is good agreement between the experimental phase line determined by ZSS and recent theory of Pierleoni et al [6]. Twenty years ago Weir, Mitchel, and Nellis [7] measured metallic electrical conductivity of MH in a reverberating shock wave experiment, finding ~2000 S/cm. By contrast MDMG calculate DC conductivity of ~15 S/cm, two orders of magnitude smaller. There is no clear explanation for these differences from earlier work.

This is a very challenging, complex experiment that relies heavily on interpretation. Their phase diagram is completely different from the one determined by ZSS. Hydrogen cannot have two different phase diagrams! Here we present an alternative model that can resolve the differences in the reported results.

We first point to more problems or inconsistencies in the paper by MDMG. In contrast to their interpretation, they probably produced MH, but their resolution was inadequate to observe the transition as a plateau in T in a heating curve, as was done by ZSS and Ohta et al [8]. MDMG pulse laser heat their sample with an approximately 5 µs long laser pulse having an unusual shape shown in Fig. 1a. The pulse has an initial ~200 ns wide peak with ~3 times the power of the remaining 5 µs, and then slowly decays. The power is sufficient to heat to 5000-6000 K in the initial peak. The phase transition to MH can be identified by a plateau in a heating curve (a heating curve is a plot of sample temperature vs laser power or vs time). Temperatures rise rapidly in laser heated samples, so plateaus in heating curves are created at very short times. This has been shown experimentally by ZSS, and in a temperature/time curve studied with an FEA by Geballe and Jeanloz [9] (Fig. 2). For the very large pulse power used by MDMG the temperature rises far above the plateau temperature observed by ZSS and would be a small "wrinkle" in the T vs. time curve (Fig.2). At MDMG's pressure of ~141 GPa, if MH was created, they would not have seen the associated plateau: they sample their data every 1000 ns and report a time resolution of 2000 ns for temperature measurements.

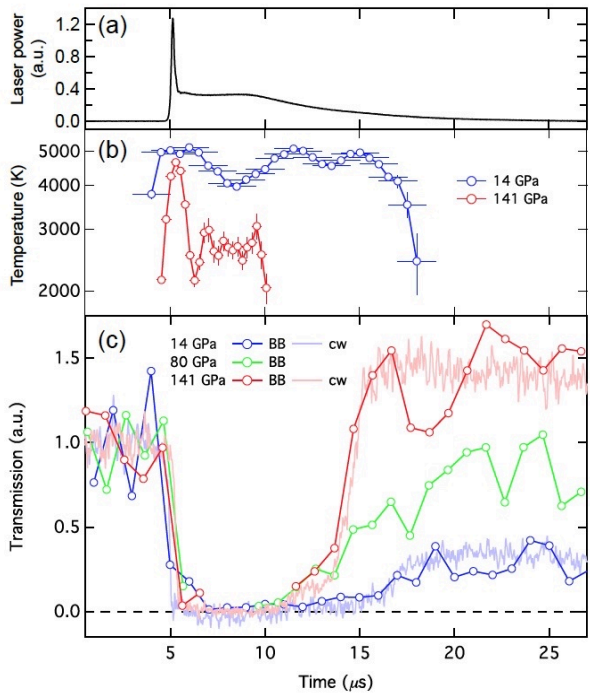

Fig. 1. Fig. S4 from MDMG. (a) shows the shaped laser pulse used to heat the hydrogen sample. (c) Transmission as a function of pulse time. The red circles or pink curve are for P=141 GPa and are consistent with zero transmission during the heating pulse. MDMG state "However a high transient extinction due to sample absorption could still be resolved."
(b) shows the measured temperature for a low pressure heating (blue points) and high pressure (red points).

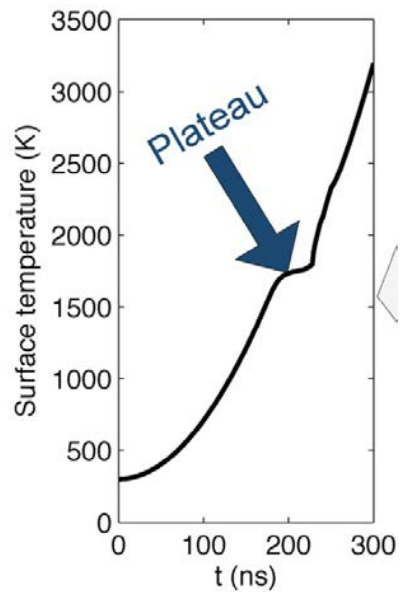

Fig. 2. A temperature/time heating curve from the FEA of Geballe and Jeanloz. This shows a plateau due to the latent heat of fusion of a 5 micron thick absorber.

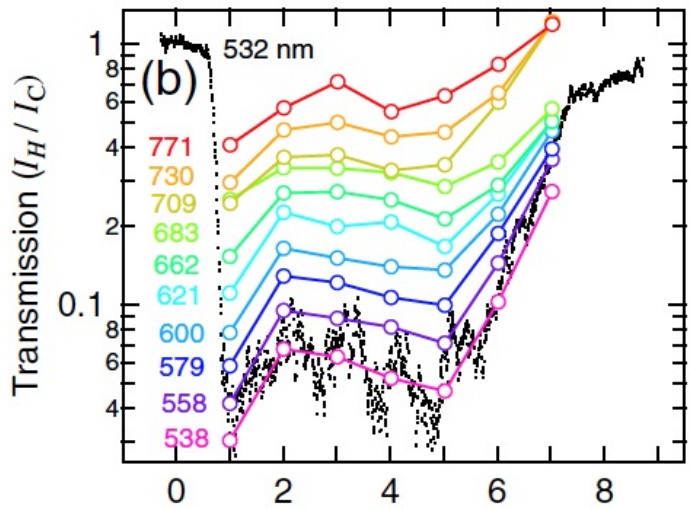

Fig. 3. The wavelength dependence of transmission for hydrogen at a pressure of 141 GPa, extracted from Fig. 2 of MDMG. The black curve with unexplained oscillations is continuous transmission of a 532 nm laser beam.

Now, consider MDMG's data. In their Fig. 2 (our Fig. 3), they show finite transmission data as function of wavelength and time. However, in their Fig. S4-c (Fig. 1) for 141 GPa and ~3000 K, their data shows <u>zero</u> transmission. Nevertheless they report weak transmission and interpret their data assuming that hydrogen is a semiconductor. A semiconductor should have a gradual increase in absorbance with rising temperature, as the conduction band is populated or narrowed. They report an abrupt increases in absorbance as temperature is increased, such as might be observed in a phase transition, but they do not discuss the basis of this behavior. They claim that their sample is strongly absorbing so that it would have a large complex part of the index of refraction, giving rise to large reflectance, yet they calculate reflectance to be less than 1%, also assuming the real part of the index to be ~1. It is known to be ~2.4 at their highest pressures in molecular hydrogen. Finally, they do not give any detail on how they determine their temperature from the gray body radiation. For this determination it is important to know the wavelength dependence of the emissivity of their hot surfaces to determine a correct transfer function. In their experiment they have three surfaces that change during their pulsing: iridium, "absorbing hydrogen", and Ir surfaces that are attacked to form $IrH_3$.

We believe that the observations by MDMG can also be explained if the hot 141 GPa sample became a thick metallic film. In this case there would be no transmittance through the hole in the iridium laser absorber, as seen in their Fig. S4-c (our Fig. 1).

How does light get to the spectrometer if the sample is metallic and non-transmitting? In spite of the observation that Fig. 4S-c (our Fig. 1) shows zero transmission, it remains to understand how light might get to the spectrometer/detector if the sample is metallic and non-

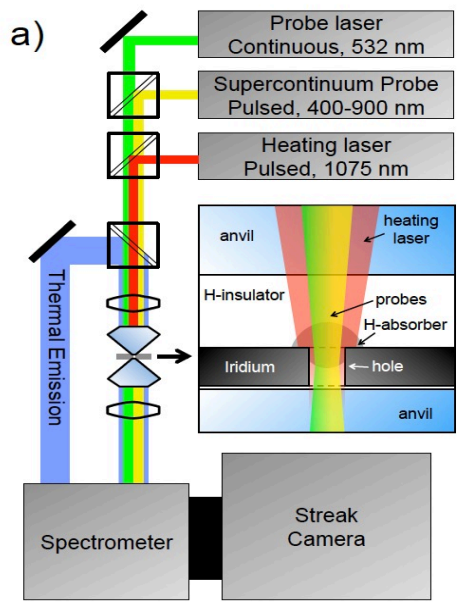

Fig. 4. Fig. S1 from MDMG, showing their experimental schematic.

transmitting, as we propose. The rendering of the sample cell in their Fig. S1 (our Fig. 4) has two problems:
- First, the sample is heated by conduction from the absorber. Therefore the hot sample should back-heat the absorber, which has a large metallic thermal conductivity. Moreover MH has a large thermal conductivity, so that energy absorbed in a thick film of MH will conduct to the absorber to maintain its temperature. Note that the absorber is insulated from the diamond by a thin layer of molecular hydrogen so that the entire absorber can heat up within a few thermal time constants, not just the surface, as shown in Fig 5.
- Second, their Fig. S1 (our Fig. 4) does not show the full geometry, as there is space between the absorber and the gasket hole (using information from their SM; see also [2]). An alternate interpretation is provided in our Fig. 5, which shows the gasket and possible pathways for light to arrive at the detector without going through the hole in the absorber.

In order to test this hypothesis, we have prepared a diamond anvil cell with a gasket having a 95 micron diameter hole. We replace their absorber with a reflecting aluminum "absorber", without a central hole, that was placed on one of the diamond culets. We used a rectangular absorber (50x60 nm) and focused an expanded laser beam (532 nm wavelength) to a spot (a few microns in diameter) on the back of the absorber. The gasket/absorber region was magnified by a factor of 10 and imaged onto a CMOS camera in the focal plane of the lens. In principle if there were no diamonds and the light was focused on the absorber, no light would get to the transmission detector. A spatial filter can pick out areas of interest in its focal plane, occupied by the camera in our setup. An image of the gasket cavity and absorber showing the light that passes through the gasket hole is shown in Fig. 6. Intense light emerges from the space between the absorber and the gasket edge. This light can be blocked with the spatial filter utilized by MDMG; however the light that appears to be coming from the backside of the absorber <u>cannot</u>! This light arises from the focused laser beam that must be multiply reflected, as indicated in Fig. 5. Note that the Tolkowsky design of a diamond [10] optimizes light entering the diamond table to reflect back out of the table; thus light can reflect from the table to the absorber and back out the table to the detector (it is this property that gives diamond its sparkle). Thus, the transmitted signal measured by MDMG and interpreted as absorption of hydrogen, may just be light, first reflected off of MH, that could not be blocked by their spatial filter.

It remains to explain the wavelength dependence of the transmission reported by MDMG in their Fig. 2. We believe that the hydrogen around the periphery of the absorber is hot, but remains molecular with a small band gap at high density. The "leaking" light passes through this region and can be responsible for the observed wavelength dependence of absorption. Thus, the model we present can resolve the differences in the phase diagrams.

We thank Ranga Dias, Ori Noked, and Bill Nellis for discussions concerning our comment. The NSF, grant DMR-1308641, the DOE Stockpile Stewardship Academic Alliance Program, grant DE-NA0001990, and NASA Earth and Space Science Fellowship Program, Award NNX14AP17H supported this research.

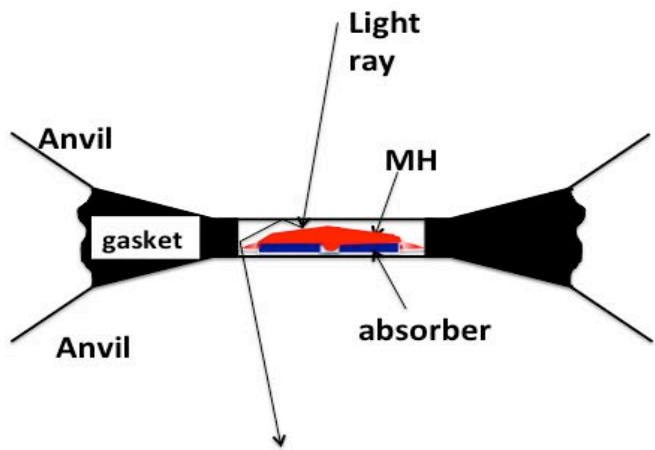

Fig. 5. An alternative schematic of MDMG's experiment, showing the open space at the periphery of the absorber. Light rays can possibly reflect off of the MH and off of the diamond culet, off of the diamond table (not shown), and back to the absorber.

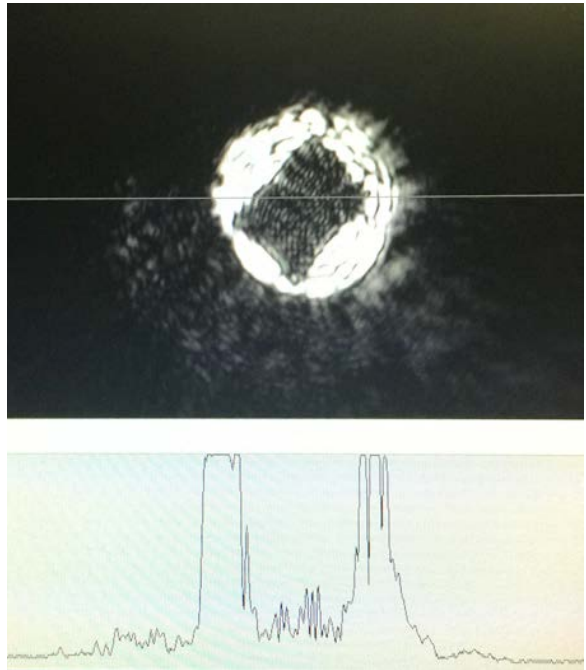

Fig. 6. Image of a DAC cell, showing the Al absorber side facing the transmission detector. This is back-lit with a focused 532 nm laser beam. The lower part of the figure shows the light intensity across the horizontal line in the upper part. It is clear that light rays can appear to be coming from the opaque side of the absorber, which cannot be spatially filtered out.